\documentclass[14pt]{article}
\usepackage{graphicx}
\usepackage{xcolor}

\begin{document}
\title{Lattice Fluid Dynamics: Thirty-five years Down the Road}
\author{Sauro Succi} 
\title{Lattice Fluid Dynamics: Thirty-five Years Down the Road}
\author{Sauro Succi $^{(1,2,3)}$\\ 
(1) Center for Life Nano-Neuro Science at La Sapienza, \\ Italian Institute of Technology,
Viale Regina Elena 295, 00161, Roma, Italy,\\
(2) Physics Department, Harvard University, \\Oxford Str 17, 02138, Cambridge, USA,\\
(3) Mechanical Engineering Department,\\ University College London,Gower Street, London WC1E 6BT (UK)   
}

\maketitle

\abstract{
We comment on the role and impact of lattice fluid dynamics 
on the general landscape of computational fluid dynamics. 
Starting from the historical development of Lattice Gas Cellular Automata, we 
move on to a cursory appraisal of the main applications of the Lattice Boltzmann 
method and challenges ahead.
}

\section{Introduction}

Computational Fluid Dynamics (CFD) has been around for as long as 
computers exist, starting with von Neumann's program to simulate the weather
on the ENIAC machine (1950's) and even earlier, with 1922 Richardson's 
description of "human" computers computing the weather by hand (he estimated that 64,000
human calculators, each calculating at a speed of 0.01 Flops/s, would 
be sufficient to predict the weather in real time
(https://www.altair.com/c2r/ws2017/weather-forecasting-gets-real-thanks-high-
performance-computing).
Leaving aside human calculators, electronic ones have made a 
spectacular ride till current days, from the few hundred Flops of ENIAC
to the current few hundred Petaflops of the Top 1 IBM-Nvidia Summit computer.
Sixteen orders of magnitude in 70 years, close to a sustained Moore's law rate
(doubling every 1.5 years)!
Amazingly, CFD has been consistently on the forefront of such spectacular ride
and continues to do so to this day. 
Finite Differences served as the tool of the trade in 60-70's, later to be replaced 
by finite-volumes and, to a lesser extent, by Finite-Elements as well on account of the need of
handling complex geometries, a must for most engineering problems. 
On a more fundamental side, spectral methods, have taken the lions's share of CFD
for the numerical study of homogeneous fluid turbulence, and keep holding 
their pole position to this day.
All of these methods respond to the same strategy: {\it Discretization of the 
continuum equations of fluid mechanics}.
That is, they start from the Navier-Stokes equations  
and represent them on a suitable grid in real or reciprocal space, sometimes both.
This is the mainstream numerics for basically any classical continuum field equations,
but in the mid 80's a qualitatively different approach was put forward by a small group
of inspired researchers in France and the USA.

\section{Lattice Gas Cellular Automata} 

In the mid 80's a very different approach was proposed: instead of discretise
the continuum fluid equations, one rather devises a fictitious particle dynamics
which would recover the continuum picture in the infrared limit: large scales
as compared to the particle mean free path \cite{FHP,WOLF1}.
At first, this sounds like a self-inflicted pain,  for it is known that
molecular dynamics can barely reach micrometric regions in space and milliseconds
in time, even on the largest supercomputers.
The key, though, is that particle are not real molecules but "effective" ones, i.e.  
they represent a large collection of real molecules, where large means basically 
the number of molecules in a single computational cell.

It turns out that if the dynamics of these fictitious particles can be designed 
in such a way as to preserve the basic mass-momentum-energy conservation laws, relinquishing 
all the hydrodynamically irrelevant molecular details, this fictitious particle
dynamics (FPD) would prove competitive against the 
mainstream discretization approach.

A spectacular example in point is the famous Lattice Gas Cellular
Automata (LGCA) first proposed by Frisch, Hasslacher and Pomeau in 1986 \cite{FHP}
with substantial early contributions by Stephen Wolfram as well \cite{WOLF1}.
In equations:
\begin{equation}
\label{LGCA}
n_i(x+c_i,t+1)-n_i(x,t) = C_i[n]
\end{equation}
where $n_i={0,1}$ is a boolean occupation number  indicating the 
absence(presence) of a particle at site $x$ and time $t$
with discrete velocity $c_i$ (vector indices relaxed for simplicity).  
The LHS of the above equation represents the free-streaming of a boolean 
particle with velocity $c_i$ from site $x$ at time $t$ to site $x_i = x+c_i$ at time $t+1$. 
The streaming is synchronous, in that all sites $x_i$ still belong to 
the uniform regular lattice.  This step is exact, in that it implies an error-free 
transfer of information from site $x$ to $x_i$, no loss in between. 
The RHS codes for the local particle collisions at site $x$ and time $t$.  

The beauty is that it does not have to conform to any detail of molecular 
interactions, but only to the main conservation laws: mass-momentum-energy 
(the latter not being part of the original formulation) as well as rotational invariance.

Remarkably, the above prescriptions can be encoded in 
Boolean form too. To be precise, the collision operator takes 
the values $C_i ={-1,0,+1}$ corresponding to annihilation, 
no-action, generation of a particle with velocity 
$c_i$ at site $x$ and time $t$.  
A typical  collision $(i,j) \to (k,l)$ is then coded
by a fourth-order boolean polynomial of the form:
$$
C_{ij \to kl} = n_i \cdot n_j \cdot \bar n_k \cdot \bar n_l
$$
where $\cdot$ is a logical AND and overbar denote logical NOT
(turning 0 into 1 and viceversa).
Note that the above term involves a four-body interaction because 
boolean particles are fermions,  hence the collision can take place only 
on condition that the final states are empty.

The major appeal of LGCA rest with its fully Boolean nature, hence 
literally round-off freedom, as well as their outstanding
amenability to parallel computing.
The expectations were high, to the point that the Washington Post 
maintained that the method should be "kept off the Soviet hands" \cite{WASH} 
(for the record, back in 1986 the Berlin wall was still standing).

Despite the tantalizing promises, the LGCA did not make to the mainstream 
CFD, for a series of conceptual and technological reasons,  including 
statistical noise,  low collisional rates
and exponential complexity of the collision rule with the
number of discrete states.

In hindsight, the main show-stopper was the latter.
Indeed the complexity of the collision operator, scales like $2^b$ for a 
LGCA with $b$ discrete states per site ($b$ stands for "bits").
It turns out that in three spatial dimensions
the minimum suitable hydrodynamic lattice is a {\it four}-dimensional Face
Centered HyperCube (FCHC), featuring $b=24$ and resulting 
in $O(2^{24} \sim 16$ millions) boolean operations at each lattice site and every time step.
Too much to compete with "standard" methods, requiring of the order 
of hundreds of floating point operations for the same task.  
In other words, LGCA proved unviable because of their exponential 
computational intensity.

Heroic efforts were put in place, mostly by the French school 
to mitigate the problem,  i.e.  minimise the number of collisions
for a given value of the Reynolds number \cite{HENON}.
Even so,  the result did not change the ultimate sentence
of computational unviability, so that practical interest 
in LGCA started to flag down in the early 90's. 

It would be a gross mistake to identify this
computational "failure" with a scientific flop.
Quite on the contrary,  LGCA was an overly
productive idea,  as it opened the way to the possibility that 
fluids can be simulated using a suitably stylized microscopic approach
rather than by discretizing the continuum equations of fluids.

Among others, this idea gave birth to the highly successful 
offspin known as Lattice Boltzmann (LB) method \cite{OUP01,OUP18,AIDUN}.

Even though in hindsight it appears "obvious" that LB could have been 
generated by direct discretization of the continuum Boltzmann kinetic equation,  
an observation that led a significant number of scientists to dismiss the LGCA
and the historical LB route, as needlessly complicated.
This is very easy, but only in hindsight: the fact remains that historically 
this opportunity was realised and seized only thanks to the LGCA work.
Till then, discrete velocity models, which predated LGCA
by two-three decades \cite{DVM}, were never meant to be used 
for fluid-dynamic purposes!

\section{Entry Lattice Boltzmann}

Be as it may, historically LB developed in the wake of LGCA. 

The basic idea of LB is to replace effective particles with corresponding
probability distribution functions (PDF), along the same line which
takes molecular dynamics to Boltzmann's kinetic theory.
On the lattice though the task is littered with deadly catches, namely
badly broken symmetries that would prevent the recovery of the correct
equations of fluids in the continuum limit. 
The Lattice Boltzmann (LB) method made its earliest chronological
appearance in 1988 \cite{MZ}, its first computationally 
viable realization being published just months later \cite{HS}.
Ever since, LB has marked a tremendous growth in methods and 
applications \cite{HJ,HSB,BSV,REC,LBGK,LBGK1,LBGK2,MRT,ELB,
HERMI,GUO,STATPHYS,THE1,THE2,THE3} 
over  an amazingly broad spectrum of problems across different regimes 
and scales of motion,  literally from astrophysical jets\cite{REZZO} all the way down to 
quark-gluon plasmas \cite{EPL15}.

\section{LB in a nutshell}

The LB equation (LBE) reads as follows:

\begin{equation}
\label{LBE}
f_i(\vec{x}+\vec{c}_i,t+1)-f_i(\vec{x},t)=
-\Omega_{ij} (f_j-f_j^{eq})(\vec{x};t) + S_i(\vec{x};t)
\end{equation}
where the lattice time step is made unit for simplicity.
In the above, $f_i(\vec{x},t) \equiv f(\vec{x},\vec{v}=\vec{c}_i;t), \;i=0,b$ 
denotes the probability of finding a particle at lattice site 
$\vec{x}$ and time $t$ with a molecular velocity $\vec{v}=\vec{c}_i$.
Here, $\vec{c}_i$ denotes a set of discrete velocities, which 
must exhibit enough symmetry to
obey the mass-momentum (energy) conservation rules, along with 
rotational invariance (isotropy).

Typical two and three dimensional lattices are shown in Fig. 1.
\begin{figure}
\centering
\label{FigLat}
\includegraphics[width=3.5cm,height=3.5cm]{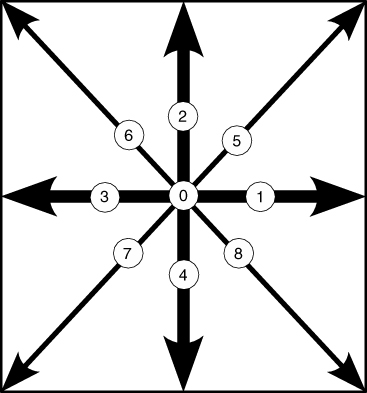}
\includegraphics[width=3.5cm,height=3.5cm]{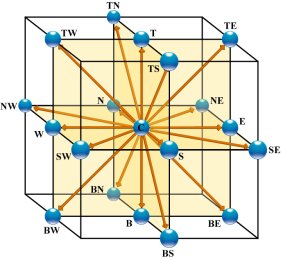}
\caption{The workhorse lattices in two and three dimensions, with
$9$ and $19$ discrete velocities, respectively.}
\end{figure}  
The left hand side of eq. (\ref{LBE}) denotes free-streaming, while 
the right hand side is the collision step, which
consists of a short-range component, driving the system towards a local
equilibrium $f_i^{eq}$ and a soft-core source of momentum $S_i(\vec{x};t)$.
The latter is in fact more general and can represent the source of any
macroscopic quantity relevant to the physics in point.  
This is a major asset of LB, one that permits to include complex mesoscale
physics often beyond the realm of continuum fluids, at a minimum cost in terms
of programming and computing overheads. 

The local equilibrium corresponds to a 
finite-order truncation of a local Maxwellian:
\begin{equation}
f_i^{eq} = w_i \rho(1 + \beta u_i + \frac{\beta^2}{2} q_i)
\end{equation}
where $\beta=1/c_s^2$, $u_i = \vec{u} \cdot \vec{c}_i$ 
and $q_i=u_i^2-u^2c_s^2$, $c_s$ and $\vec{u}$ being the 
lattice sound speed and the local flow velocity, respectively.
In the above $w_i$ is a set of weights normalized to unity, the
discrete analogue of the absolute Maxwellian in continuum velocity space.
The truncation is an unavoidable consequence of lattice 
discreteness, which permits to recover Galilean invariance
only to a finite order in the Mach number $Ma=|\vec{u}|/c_s$.
The relevant hydrodynamic quantities are computed as simple
{\it linear and local} combinations of the discrete distributions, namely:
\begin{equation}
\rho(\vec{x};t) = \sum_i f_i(\vec{x};t),\;\;\;
\rho \vec{u}(\vec{x};t) = \sum_i \vec{c}_i f_i(\vec{x};t)
\end{equation}
where $\rho$ is the fluid density.
Higher order moments deliver the fluid pressure
and stress tensor in the form of {\it linear and local} 
combinations of the discrete populations, which 
proves very convenient for simulation purposes.
Formally, The LB eq. (\ref{LBE}) is nothing but
a set of finite-difference equations, and yet one with
great power inside. 
This power stems mainly from four basic ingredients, namely:
{\it
i) Exact free-streaming, ii) Local lattice equilibria, 
iii) Tunable relaxation matrix, iv) Flexible external source,
}
Before moving on to these items, we hasten to add that the eq. (\ref{LBE}) can 
be shown to converge to the (quasi-incompressible) Navier-Stokes 
equations in the usual limit of small
Knudsen numbers, $Kn=\lambda/L \ll 1$, i.e. small mean 
free path versus the typical scale of variation of hydrodynamic quantities.
This is also a statement of weak departure from local equilibrium.  
Technically, this entails a Taylor expansion in the lattice 
time step, as combined with a double expansion in low Knudsen and Mach number.
The tool of the trade is the Chapman-Enskog asymptotics \cite{CHAP}. 

\section{Mainstream LB applications}

As mentioned above,  at the time of this writing LB is a
massive bibliographic presence in fluid dynamics and most allied
fields, particularly soft flowing matter \cite{OUP18} and the physics/biology
interface \cite{OUP22}. 
It would be utterly impossible to cover this vast ground in
any single paper, hence here we shall just outline the main
applications in very broad strokes. 

\subsection{Macroscopic flows}

Historically, Lattice Boltzmann was developed as a computational
alternative to the discretization of the Navier-Stokes equations
of continuum fluid dynamics, the main target being high-Reynolds
turbulent flows in complex geometries \cite{SCI03}.
The initial hopes were that LB would offer a better resolution of 
the  near-grid scales or even provide a natural subgrid model via 
the extra degrees of freedom of the kinetic representation 
(then dubbed "ghost modes") versus the hydrodynamic one \cite{JPA}.
\begin{figure}
    \centering
    \includegraphics[scale=0.75]{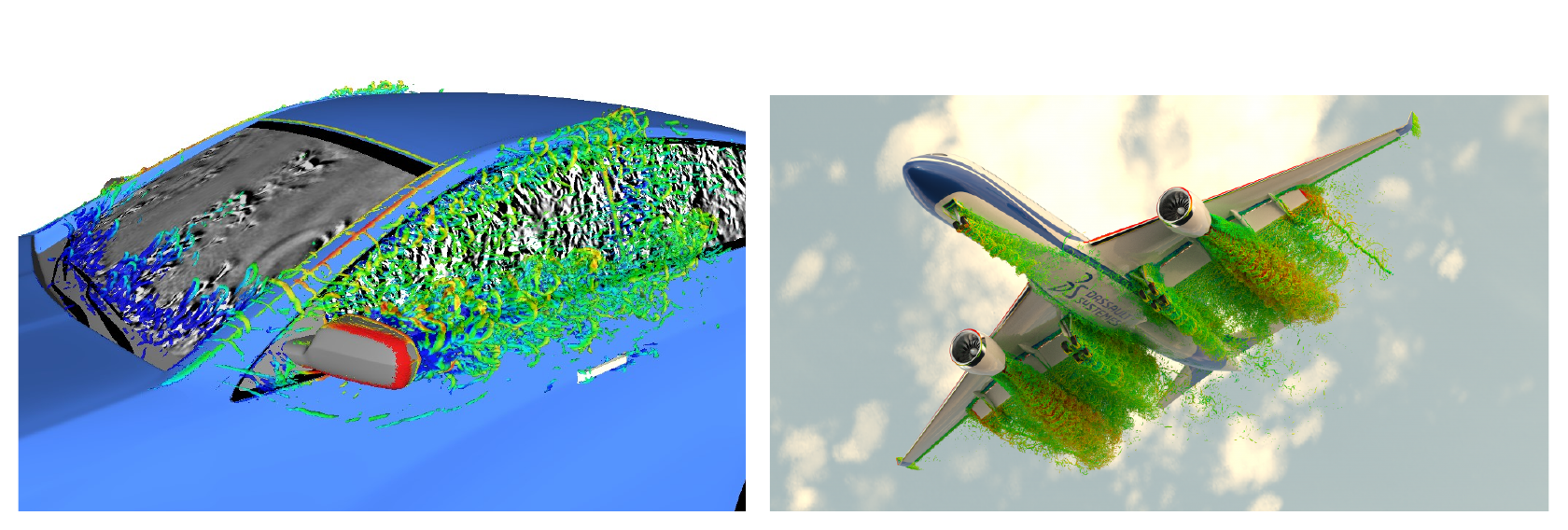}
    \caption{ Two examples of fully-turbulent lattice Boltzmann simulations for automotive and aeronautical applications. (Courtesy of Dassault Systemes) }
    \label{fig:turbo}
\end{figure}
This turned out not to be the case, but prepared the ground for subsequent developments which have met with significant success, as witnessed by existence of a number 
of open source \cite{PALAB} and commercial codes, particularly
POWERFLOW, developed by EXA Corporation and recently acquired by Dassault Systemes \cite{POWER}. The academic side has also witnessed remarkable progress, mostly but not exclusively 
in connection, with the development of the entropic 
method \cite{TURBOELB} and Large-Eddy simulations \cite{SAGA1,SAGA2,SANKYA}.
More recently, LB has also been used to study turbulent flows with suspended 
bodies, a topic of great relevance for energy and environment \cite{TOSCHI}.
Another successful application of LB to macro-hydrodynamics
are flows in porous media,  with accompanying heterogeneous
chemical reactions \cite{LBPOR1,LBPOR2}, 
which draw major benefits from the LB ability to deal 
efficiently with grossly irregular geometries.

\subsection{Multiphase and colloidal flows}

Two areas where LB has made a real difference are 
multiphase/multi-component flows and flows with suspended bodies \cite{KRUG}.
In the former case LB offers the major benefit of simplicity: interfaces need no explicit
tracking but emerge spontaneously, informed by the corresponding mesoscale forces
which are implemented as soft source terms. This benefits comes at the prize
of several limitations, most of which have been however significantly mitigated
in the course of time \cite{MU1,MU2,MU3}.
\begin{figure}
    \centering
    \includegraphics[scale=0.8]{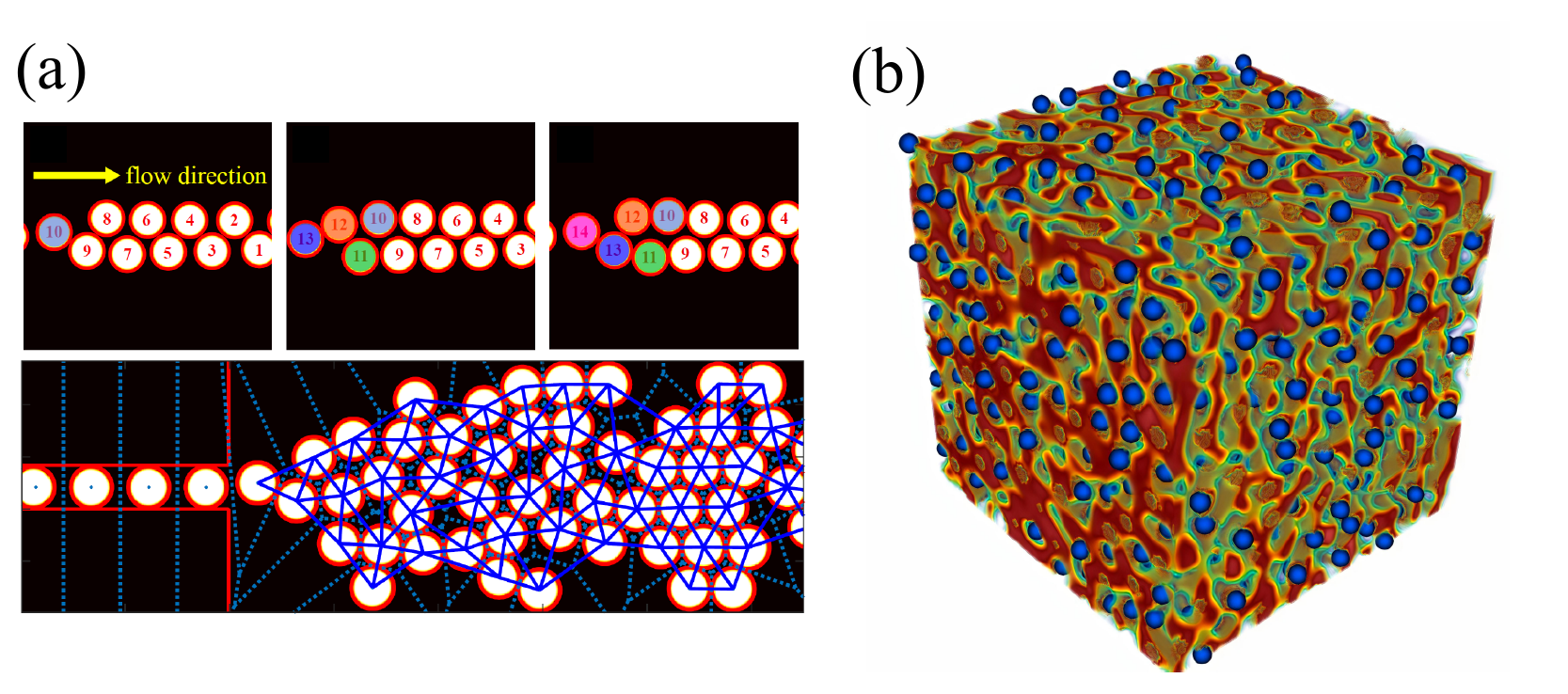}
    \caption{(a) Spontaneous ordering of droplets in a microfluidic channel
     (\cite{monte2021soft}). (b) Large-scale simulation of a colloidal flow (\cite{bonaccorso2021lbcuda})}
    \label{fig: colloids}
\end{figure}
Many LB variants exist today which have found massive use especially in 
the area of microfluidics (see below).
Among these, a very fruitful option is offered by 
the  {\it Color Gradient} technique \cite{gunstensen1991lattice}, in which
the idea is to add an explicit anti-diffusive flux sending particles of each species 
uphill along their density gradient instead of against it, thereby promoting the formation 
of an interface against the coalescing effect of surface tension. 
The stress-jump condition across fluid interfaces can be further augmented with 
an immersed-like force modeling the repulsive effect generated by a surfactant 
solution absorbed onto the drop interfaces. 
This contribution can be added to the collision operator via a suitable forcing term as proposed in \cite{monteJFM2019,monte_prf}.
The above approach has been shown to correctly capture highly non trivial 
many-droplet configurations,  like dense emulsions and foams.
As an example, the extended multi-component model has been shown to reproduce 
the formation of ordered droplets clusters in microfluidic channels \cite{monte2021soft}.
As shown in figure \ref{fig: colloids}(a), the droplets, continuously injected within the main channel, undergo a  spontaneous ordering into hexagonal clusters, which is due to a subtle competition between 
local, short-range, repulsive interactions (i.e., the near-contact forces) and the surface tension. 
Another major field of application are complex flows with immersed bodies, both
rigid and deformable, in which LB is typically coupled to other methodologies
for fluid structure interactions \cite{KRUG,IBLB}. 
To this regard, recent Multi-GPU state-of-the art implementations of  LB codes for
multi-component colloidal flows \cite{bonaccorso2021lbcuda} have been shown to  
deliver remarkable performances, up to $200$ GLUPS (Giga
Lattice Updates Per Second on a cluster made of $512$ A100 NVIDIA cards),
on computational grids with several billions of lattice points ( see figure \ref{fig: colloids}(b). 
These results open attractive
prospects for the computational design of new materials based on colloidal particles.

\subsection{Micro and nano-hydrodynamics}

In the early 1990's LB was applied exclusively to
macroscopic fluid problems, and even with a vengeance, since some
prominent researchers vetoed its use for anything but this \cite{LUO}.
Yet, precisely in those years, a few groups offered 
numerical evidence that the LB scheme can 
handle moderately non-equilibrium flows beyond the strict
realm of hydrodynamics \cite{HOL,AKBC,monte2015}.
This, together with a boost of activity in the area of 
non-ideal, multiphase and multicomponent fluids, has led to a major 
mainstream of LB applications, mostly in soft-matter
and microfluidics 
\cite{monteJFM2019,SOFT2,SOFT3,SOFT4,CHOPA,SUGA,BENZI06,tiribocchiNatComm2021}. 

The key ingredients for this transition from the macro
to the micro-hydrodynamic levels turned out to be the 
introduction of higher order lattices, securing the compliance  with generalized 
hydrodynamics beyond the Navier-Stokes description, as 
well as the development of suitable boundary conditions, capable 
of dealing with non-equilibrium effects associated with large gradients at solid walls
\cite{AKBC,THAMPI}.  
Very recently, the LB has also been successfully coupled to 
off-lattice, particle models to develop novel classes of fully mesoscale hybrid approaches 
capable of capturing the physics of fluids at the micro‐ and nanoscales whenever 
a continuum representation of the fluid falls short of providing the complete physical information. 
In addition, LBM has also been coupled to  Direct Simulation Monte Carlo (DSMC) in 
view of efficiently simulating isothermal flows characterized by variable rarefaction effects \cite{distasoJCS}.

Another major mainstream of current LB research is the study of
flows with suspended bodies, both rigid and deformable ones. 
This was originated by pioneering work on colloidal nanoflows, i.e.
rigid spheres flowing in a LB solvent, which was sparked by
the development of the so-called fluctuating Lattice Boltzmann, taking 
in due account the statistical fluctuations which characterize hydrodynamic 
behavior at the nanoscale \cite{LADD}.

The original fluctuating LB (FLB) scheme for colloidal flows has 
spawn many subsequent developments, including biopolymers and 
deformable objects of biological interest, such as membranes and cells.
Methodologically, this was made possible by the fruitful merge between
LB and the Immersed Boundary Method \cite{IBLB}.
Details of this  highly technical and utterly important subject can be found in \cite{KRUG}.

\section{Quantum-relativistic fluids}

Even though the mainstream of LB research remains 
hard-rooted it classical physics, it should be mentioned that 
the LB scheme is amenable to remarkable generalizations 
in the direction of both quantum and relativistic mechanics.
The former possibility was recognized as early as 
1993, building on a formal analogy between the Boltzmann equations 
and the Dirac equation of relativistic quantum mechanics \cite{QLB,DELLAR,FILLI}.
Incidentally, the corresponding quantum LB (QLB) scheme was later shown
to represent a quantum random walk, as per the seminal Ahronov et al
paper, which appeared just a couple of months ahead of QLB.
The QLB was also shown to be amenable to quantum computing \cite{YEPEZ}.  
   
The extension of the LB scheme to relativistic fluids came much later,  as it dates 
of 2010 \cite{RLB10,RLB11} and it has undergone major progress in the
last decade \cite{GABBA1}.
The relativistic LB requires some non-trivial technical manipulation due
to the fact that, unlike the Newtonian case, the relativistic energy 
of massive particles is  an irrational function of 
the particle momentum, i.e.
\begin{equation}
E(p) = \sqrt {m^2 c^4 + p^2 c^2} 
\end{equation}   
This makes the task of developing discrete-velocities schemes 
less straightforward than in the non-relativistic case, $E \ll mc^2$,
{\bf because one can no longer rely on orthogonal polynomials in 
the continuum, such as Hermite's.  However, suitable discrete 
orthogonalization techniques have been developed over the years,
which have permitted to realize fairly efficient relativistic LB schemes
despite the large number of discrete speeds involved
\cite{GABBA2}.
}

This has permitted the simulation of relativistic flows, such as electron 
transport in graphene and transport phenomena in quark gluon plasmas.
Amazingly, similar schemes have also been used for the simulation
of cosmic neutrinos in astrophysical flows \cite{REZZO}.
Relativistic hydrodynamics is still a small niche as compared to 
the massive activity in non-relativistic fluids.
{\bf However,  due to the mounting interest in relativistic fluids 
at the crossroads between condensed matter, 
high-energy physics and gravity \cite{POLISON},  it is reasonable to expect
that relativistic LB schemes may gain significant momentum for the years to come}.

\section{Towards exascale LB}

A common thread to all the LB applications is their
outstanding amenability to parallel computing, across 
virtually any kind of architecture (provided hard and ingenuous implementation work
is spent on the task).
Remarkable examples abound in the literature 
\cite{PALAB,MUPHY,BER10,BER13,HEMELB,WALB1,EXA17,EXA19,WALBER}, 
but here I shall just refer to two instances 
I am directly familiar with.
\begin{figure}
\label{FigLBX}
\centering
\includegraphics[width=7cm,height=5cm]{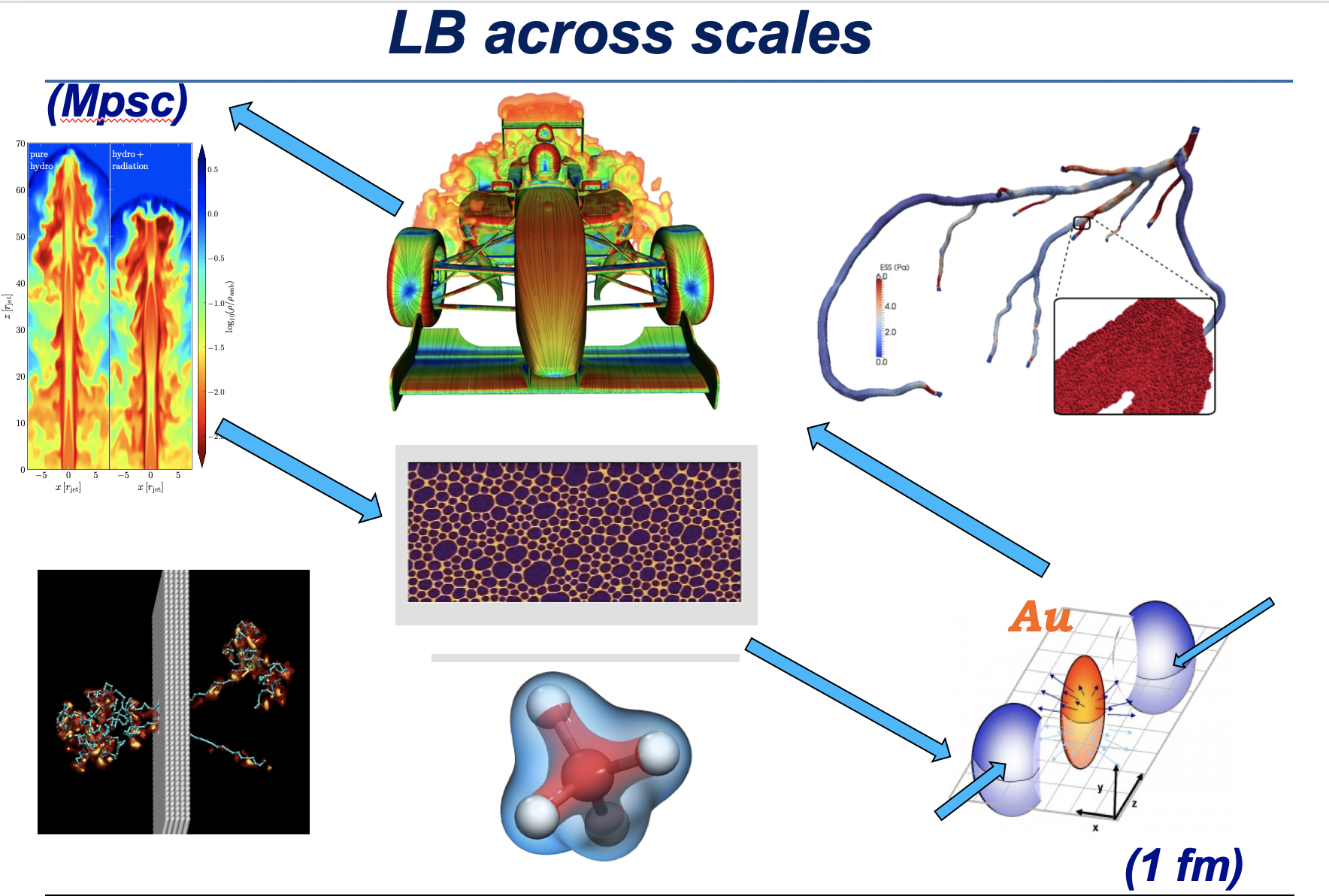}
\caption{LB across scales: from above galaxies all the way down inside 
the proton! Top row: astrorelativistic jets resulting 
from binary neutron star mergers, flow around a racing car, blodd flow in human
arteries at blood cell resolution.  Middle: a wet foam. 
Bottom row: double-stranded biopolymer trasnslocating through a membrane, 
The electronic cloud around the methane molecule, the quark-gluon plasma
produced by heavy ion collisions.
All of these flows, covering from megaparsecs to femtometers (37 decades), 
made use of different variants of the LB method.
}
\end{figure}  
First, the use LB as a fast water solvent solver 
for the study of protein crowding within the cell, which reached in excess 
of 20 Pflops on the Titan supercomputer, but many outstanding
performances have been obtained in other massively parallel 
applications at all scales of motion \cite{BER13}.
As a recent and rather spectacular example, I wish to
quote the full-scale simulation of the
deep-sea sponge Euplectella Aspergillum, with a fully realistic description of 
the highly complex geometrical porous structure at a 
10-micron resolution, all  the way 
up to the centimeters of the full sponge \cite{NAT21}.
It is fair to say that such kind of study would have been 
impossible without LB.
The above applications witness the ability of suitably optimized LB codes
to achieve Petascale performance on highly non-trivial fluid problems.

The LB assets, exact free streaming, machine-accurate conservative collisions and
flexibility towards inclusion of mesoscale physics are expected to carry over from
the Peta to the Exascale. 

However,taking LB towards the exascale will face the usual challenges common
to most numerical methods, namely the need of minimizing the cost of
accessing data \cite{DATACC}, offsetting communication overheads by overlapping communication
and computation and, a possibility unexplored so far to the best of this author's
knowledge, develop fault-tolerant LB schemes. 
{\bf All of the above,  by taking into account the increasing constraints on power consumption, for
which machine learning might prove a valuable ally, typically by optimizing the 
reconstruction of fine-grained information from coarse-grain simulations}.

\section{Towards LB Quantum Computing}

Quantum computing holds great promises to become the leading simulation
technology for the future.  To make the promise come true, however, several 
obstacles need to be circumvented,  on both hardware and software fronts.
As to latter,  one has first to identify the class of problems which can be  
formulated in terms of quantum computing algorithms.
To date,  the most likely candidates for such a paradigm shift are mostly in the
area of quantum chemistry and material science; understandably so,  since quantum
computing is obviously most suited to quantum systems \cite{FEY}.
In this respect, fluid dynamics faces with two major issues: it is 
strongly nonlinear and dissipative.
Indeed,  quantum computing for fluids is still in its infancy \cite{GAITA,CHILD}, even though
interesting efforts to circumvent the two major barriers have started to  emerge in the recent years.
In the following we just sketch the basi ideas. 

\subsection{Handling nonlinearity}

As to nonlinearity, an interesting possibility is to advocate 
Carleman linearization,  which consists in trading non-linearity 
for linearity in extra-dimensions. 
 
A simple example says it best.

Consider the logistic equation:
\begin{equation}
\label{LOGI}
\frac{dx}{dt} = x(1-x)
\end{equation}
with the initial condition $x(0)=1/2$.

The quadratic non-linearity can be formally disposed of by defining two 
independent variables $x_1 \equiv x$ and $x_2 \equiv x^2$,  so that the
quadratic logistic equation becomes formally linear, i.e.:
\begin{equation}
\label{CARLE1}
\frac{dx_1}{dt} = x_1-x_2
\end{equation}
This is obviously open and requires an additional equation for $x_2$.
By differentiating $dx^2/dt = 2x dx/dt = 2x^2(1-x)$ and defining $x_3 \equiv x^3$,
we obtain
\begin{equation}
\label{CARLE2}
\frac{dx_2}{dt} = 2x_2-2x_3
\end{equation}
This is still open ,  as it requires an additional equation for $x_3$.
It is therefore apparent that teh Carleman linearization 
is confronted  with an endless hierarchy which
needs to be truncated at some stage.

The first order truncation is $x_2=0$, which delivers $x(t) = e^t$,  just the initial
stage of the exact solution of the logistic equation, $x(t)=e^t/(1+e^t)$.
The second order truncation is $x_3=0$,  delivers a 
$2 \times 2$ linear system, whose solution is readily checked to be
$x(t)=e^t(1-e^t)$,  still not quite exact but closer than the previous one.
The name of the game is quite clear: by truncating the Carleman series at level $N$
we obtain a $N \times N$ system of first order linear ODEs, whose solution captures
the $N$-th order expansion of the denominator in the exact solution.
Clearly, the exact solution, meaning by this a solution which holds exactly 
and at arbitrarily long times, can only be recovered in the limit $N \to \infty$.
Hence the Carleman linearization buys linearity,  but at the price
of an inevitable error which grows in time.
Preliminary results indicate that the truncation error decays fast with the 
order of truncation \cite{WAEL}, but much more work is required to put 
this observation on a solid basis for the actual LB equation.
This is certainly a most intersting direction for future research.

\subsection{Handling Dissipation}

Several techniques are available to handle dissipative effects within a conservative
(Hamiltonian) formulation.  One possibility is to augment the original dissipative system
with one (or more) extra-variables, call it reservoir,  absorbing the dissipated energy
so that the global system, original plus reservoir, is by construction conservative.

Another possibility is to represent dissipative operators
(the collision operator in the case of LB) as the weighted sum of two unitaries.
Let $C_{ij}$ the Carleman-linearized collision matrix associated with the LB scheme,  one writes
\begin{equation}
C_{ij} = U_{ij} + \alpha V_{ij}
\end{equation} 
where  $U$ and $V$ are two unitaries and $\alpha$ a numerical factor between $0$ and $1$.
The price for splitting the Carleman matrix into the sum of two unitaries is that
the quantum update may eventually fail, with a probability which depends on $\alpha$.
For the case of linear advection-diffusion problems, it has been shown that 
the loss of efficiency can be contained to some 10 percent on a wide range of
$\alpha's$ \cite{MEZZA}.
Whether the same is true for the full fluid equations is a totally open question
at the time of this writing, and again an interesting topic for future research.


\section{Summary: Whither LB?}

As we have been illustrating in this short review, over the last three
decades, the LB has made proof of an amazingly flexibility, with 
a wide spectrum of applications covering fluid motion across an impressively
broad range of scales, literally from inside protons to the outer Universe 
(see figure 4)!
The impact on CFD and allied fields is massive, as reflected by its bibliographic indicators.
But even leaving aside from bibliometrics, it is fair to say that LB has greatly facilitated
the inspection of highly complex flowing states of matter which would be very hard 
to analyze with other methods, if possible at all.
The main asset behind this success is always the same: information  
travels along straight characteristics and the streaming step is 
literally exact (zero-roundoff). 
With the due amount of work, far from trivial
in the case of complex geometries, this permits to achieve high parallel 
efficiencies also in the presence of fairly complex geometries and strong
nonlinearities. In the end, this is what makes LB such a powerful tool for
the study of complex states of flowing matter across regimes and scales.
This said, many challenges still remain.
Among others,  how to control lattice artifacts in high-Reynolds 
fluid flows with immersed bodies; how to enhance the stability
of LB schemes in the presence of substantial heat transfer and
compressibility effects; how to handle large density contrasts
in multiphase flows. How to represent large nanoscale fluctuations
without hampering numerical stability.
From a more technical,  but no less important, perspective, how to 
make LB compliant with the forthcoming exascale architectures, which 
requires  a serious effort in the direction of minimizing the 
cost of accessing data \cite{EXA17,EXA19,DATACC}. 
And as mentioned above, how to make LB compliant with the (hopefully)
forthcoming quantum computing architectures.
Summarizing,  the original Lattice Gas idea of solving fluid dynamics by means of 
fictitious molecules instead of discretizing continuum equation did not work in its
original and most radical (boolean) form. However, it has proved exceedingly 
fruitful from the conceptual standpoint, by providing the stepping stone 
for a new class of mesoscale methods and particularly the Lattice Boltzmann method.  
In hindsight, LB could have been derived as a discrete velocity 
version of the continuum Boltzmann equation,  with no reference to any microscale dynamics.
However, this does not change the fact that, without the LGCA inspiration, LB would
have been discovered much later, if at all.


\section{Acknowledgments}
This research has received funding from the European Research
Council under the Horizon 2020 Programme Grant Agreement n. 739964 ("COPMAT").
The author is indebted to very many colleagues and friends in Italy and around the
world, too many to mention without taking the chance of embarrassing omissions.
Here I only wish to thank Andrea Montessori and Mihir Durve for their
generous help in putting this manuscript together.







%
%



\end{document}